\documentclass[aps,prd,
floatfix,preprintnumbers,superscriptaddress,nofootinbib,nameinlink,capitalise]{revtex4-2}

\usepackage[left=2.54cm, right=2.54cm, top=2.54cm, bottom=2.54cm]{geometry}
\usepackage{palatino}
\usepackage{amsmath, amssymb}
\usepackage{graphicx}
\usepackage{array, booktabs, multirow}
\usepackage{float}
\usepackage{colortbl}  
\usepackage[table]{xcolor}  
\usepackage{geometry}
\usepackage{amsmath}
\usepackage{hyperref}
\usepackage{subfig}
\usepackage{placeins}
\usepackage{verbatim}
\usepackage{color}
\newcommand{\bmat}{\left(\begin{array}}
\newcommand{\emat}{\end{array}\right)}
\newcommand{\be}{\begin{equation}}
\newcommand{\ee}{\end{equation}}
\newcommand{\bea}{\begin{eqnarray}}
\newcommand{\eea}{\end{eqnarray}}

\def\D{\Delta}

\begin{document}

\title{Tribrid Inflation, Type II Leptogenesis, and Observable Gravitational Waves in $SU(3)_c \times SU(2)_L \times SU(2)_R \times U(1)_{B-L}$  }

\author{Waqas Ahmed}
\email{waqasmit@hbpu.edu.cn}
\affiliation{Center for Fundamental Physics and School of Mathematics and Physics, Hubei Polytechnic University, Huangshi 435003, China}

\author{Saleh O. Allehabi}
\email{s.allehabi@iu.edu.sa}
\affiliation{Department of Physics, Faculty of Science, Islamic University of Madinah, Madinah 42351, Saudi Arabia}

\author{Farishta Israr}
\email{farishtaisrar@yahoo.com}
\affiliation{Department of Physics, Quaid-i-Azam University, Islamabad 45320, Pakistan}

\author{Mansoor Ur Rehman}
\email{m.rehman@iu.edu.sa}
\affiliation{Department of Physics, Faculty of Science, Islamic University of Madinah, Madinah 42351, Saudi Arabia}

\noaffiliation


\begin{abstract}
We present a concrete realization of tribrid inflation within the framework of the gauge group $SU(3)_c \times SU(2)_L \times SU(2)_R \times U(1)_{B-L}$. In this model, inflation is driven by the neutral components of left-handed Higgs triplets, which also play a central role in generating the observed baryon asymmetry of the universe via non-thermal leptogenesis. Tiny neutrino masses arise naturally through a type-II seesaw mechanism, facilitated by right-handed triplet fields. Supergravity corrections, stemming from the leading-order terms in a non-minimal K\"ahler potential, are essential in bringing the predictions of the scalar spectral index $n_s$ into agreement with the latest cosmological data, including the Atacama Cosmology Telescope Data Release 6, Planck 2018, and LB-BK18. A key feature of this model is the existence of a viable parameter space that predicts potentially observable primordial gravitational waves, with a tensor-to-scalar ratio $r \lesssim 0.005$, placing it within reach of upcoming experiments.
\end{abstract}

\maketitle

\section{Introduction}
Cosmic inflation provides an elegant solution to several long-standing problems of the standard Big Bang cosmology, including the horizon, flatness, and monopole problems~\cite{Guth:1980zm,Linde:1981mu,Starobinsky:1980te}, while simultaneously generating the primordial density perturbations observed in the cosmic microwave background (CMB). Among the diverse inflationary scenarios, hybrid inflation~\cite{Linde:1993cn,Dvali:1994ms,Copeland:1994vg,Senoguz:2004vu, Rehman:2009wv, Rehman:2009nq,Rehman:2009yj, Ahmed:2022rwy, Ahmed:2022dhc, Ahmed:2024rdd} and its variants, including smooth~\cite{Lazarides:1995vr,urRehman:2006hu,Rehman:2012gd,Rehman:2014rpa,Ahmed:2022vlc,Zubair:2024quc, Okada:2025lpl}, shifted~\cite{Jeannerot:2000sv,Kyae:2005fi,Khalil:2010cp,Lazarides:2020zof,Civiletti:2011qg,Lazarides:2020zof,Ahmed:2022thr,Ahmed:2022wed,Afzal:2023cyp}, and tribrid inflation~\cite{Antusch:2004hd,Antusch:2009vg,Antusch:2012jc,Antusch:2012bp} stand out for their natural integration with particle physics frameworks, particularly supersymmetric grand unified theories (GUTs). Unlike conventional models where the inflaton is introduced as a gauge singlet, tribrid inflation offers a more natural embedding by identifying the inflaton with matter or Higgs fields that already play important roles in particle physics. In this context, the sneutrino (the scalar superpartner of the right-handed neutrino) has been extensively studied as a matter-field inflaton in gauge-singlet frameworks~\cite{Antusch:2004hd,Antusch:2006gh,Antusch:2010mv,Masoud:2019gxx}. However, alternative realizations with gauge non-singlet sneutrino inflatons have also been proposed~\cite{Antusch:2010va,Masoud:2021prr}. These scenarios are particularly compelling, as they provide a unified framework capable of addressing inflation, neutrino mass generation, and leptogenesis simultaneously. 
For additional realizations involving gauge non-singlet fields, see~\cite{Antusch:2013toa}.

In this study, we present a realization of tribrid inflation within a left-right symmetric model governed by the gauge group:
\begin{equation}
G_{3221} \equiv SU(3)_c \times SU(2)_L \times SU(2)_R \times U(1)_{B-L}.
\end{equation}
This symmetry structure is a well-motivated extension of the Standard Model~\cite{Mohapatra:1974gc,Senjanovic:1975rk}, restoring parity at high energies and offering a natural origin for neutrino masses via the type-II seesaw mechanism~\cite{Magg:1979tz,Schechter:1980gr}. Within our framework,   the Higgs triplets play multiple crucial roles: they drive inflation through their couplings in the superpotential, acquire vacuum expectation values to break the left-right symmetry, and generate tiny neutrino masses through the type-II seesaw mechanism. Moreover, their decays after inflation can produce the lepton asymmetry required for baryogenesis, creating a direct link between inflationary dynamics and the matter-antimatter asymmetry of the universe~\cite{ParticleDataGroup:2018ovx}.

In constructing the inflationary potential, we incorporate supergravity corrections arising from a power-law form of non-minimal K\"ahler potential, which significantly impact the scalar field dynamics. These corrections help stabilize the inflationary trajectory, suppress higher-order contributions, including loop-induced terms and shape the inflaton potential in a way that yields observationally consistent predictions. In particular, the model predicts a scalar spectral index of \( n_s \approx 0.973 \) and a tensor-to-scalar ratio \( r \sim 10^{-2}-10^{-5} \), in agreement with the latest bounds from  Atacama Cosmology Telescope (ACT) data~\cite{ACT:2025tim}, including  Planck 2018 and LB-BK1. These values lie within the expected sensitivity of future cosmic microwave background experiments such as LiteBIRD~\cite{LiteBIRD:2022cnt,LiteBIRD:2024wix}, CMB-S4~\cite{CMB-S4:2020lpa}, and the Simons Observatory~\cite{SimonsObservatory:2018koc}.

Post-inflationary dynamics in our setup are equally rich and predictive. The decay of the Higgs triplets gives rise to non-thermal type-II leptogenesis~\cite{Hambye:2000ui,Antusch:2004xy,Sahu:2004ny,Hallgren:2007nq}, generating the observed baryon asymmetry via sphaleron-mediated transitions~\cite{Kuzmin:1985mm,Arnold:1987mh}. Importantly, the reheating temperature remains low enough to avoid the gravitino overproduction problem—a major concern in supersymmetric cosmology—without compromising the efficiency of leptogenesis~\cite{Kawasaki:2017bqm}.

This paper is organized as follows. In Section II, we introduce the supersymmetric left-right symmetric model, outlining its gauge symmetry, field content, and superpotential structure. Section III is devoted to the construction of the global scalar potential, highlighting the tribrid inflation setup and the role of the sneutrino or triplet Higgs as the inflaton. In Section IV, we incorporate supergravity corrections, emphasizing how a carefully chosen Kähler potential helps stabilize the inflationary trajectory and suppress higher-order corrections, including potentially  loop corrections. Section V discusses the inflationary dynamics in the slow-roll regime and derives analytical expressions for the key observables. In Section VI, we present a detailed numerical analysis, exploring viable parameter space regions consistent with current CMB constraints. Section VII  and VIII addresses post-inflationary physics, including reheating, non-thermal leptogenesis, and the generation of light neutrino masses via the Type-II seesaw mechanism. Finally, Section IX summarizes our main findings and outlines their implications for cosmology and particle physics.

\section{Matter and Higgs Content}

The minimal realization of the left-right symmetric model is based on the gauge group $G_{3221}$~\cite{Pati:1974yy, Mohapatra:1974hk, Senjanovic:1975rk, Magg:1979tz}, which naturally restores parity symmetry at high energies. The transformation properties of the matter superfields under this gauge group are given by 
\begin{align*}
Q &= (3,2,1,\tfrac{1}{3}),  && Q^c = (\bar{3},1,2,-\tfrac{1}{3}), \\
L &= (1,2,1,-1),   &&  L^c = (1,1,2,1),
\end{align*}
where $Q$ and $L$ denote the left-handed quark and lepton doublets, while $Q^c$ and $L^c$ are their right-handed counterparts.
The Higgs sector is extended to include bidoublet and triplet superfields to achieve spontaneous symmetry breaking and incorporate mechanisms for neutrino mass generation and baryogenesis via leptogenesis. Their representations under $G_{3221}$ are:
\begin{align*}
H &=  (1,2,2,0), \\
\Delta^a_L &= (1,3,1,2),  && \bar{\Delta}^a_L = (1,3,1,-2),  \\
\Delta_R &= (1,1,3,-2),  && \bar{\Delta}_R = (1,1,3,2).
\end{align*}
Here, the index $a = 1,2$ denotes two generations of left-handed Higgs triplet superfields. For simplicity, only one pair of right-handed triplets, $(\Delta_R, \bar{\Delta}_R)$, is included. The presence of two left-handed triplet pairs is essential for inducing CP violation via their mixing, which is an important feature for realizing non-thermal type-II leptogenesis. This leads to a slight asymmetry in the Higgs sector, which does not affect the core symmetry structure or the consistency of the model.

The effective superpotential governing the interactions among the superfields is constructed with both renormalizable and non-renormalizable terms as:
\begin{align}
W ={} & S \left[ \frac{(\Delta_R \bar{\Delta}_R)^2}{M_S^2} - M_X^2 \right]
+ \frac{\lambda_{ab}}{M_S} \Delta_L^a \bar{\Delta}_L^b \Delta_R \bar{\Delta}_R
+ \frac{\gamma_a}{M_S} H H \bar{\Delta}_L^a \bar{\Delta}_R \nonumber \\
& + f_a L L \Delta_L^a + \beta L_c L_c \Delta_R
+ Y^l L L_c H + Y^q Q Q_c H + \mu H H , \label{eq:superpotential}
\end{align}
where $a,b = 1,2$, and the $SU(2)$, color, and flavor indices have been omitted for clarity. The constants $M_X$ and $M_S$ represent a superheavy mass scale and a cutoff scale, respectively. To simplify the structure, the coupling matrix $\lambda_{ab}$ is taken to be real and diagonal, i.e., $\lambda_{ab} = \delta_{ab} \lambda_a$.  The couplings $\gamma_a$, $f_a$ and $\beta$ are generally complex, while $Y^l$ and $Y^q$ denote the Yukawa couplings for leptons and quarks. The $\mu$-term is assumed to be of the order of the electroweak scale. A natural realization of the $\mu$ term, along with a related hybrid inflation scenario, can be found in a similar model based on the same gauge symmetry but with a different pattern of symmetry breaking, as discussed in \cite{Ahmad:2025dds}.

To control the structure of the superpotential and suppress undesirable terms, two additional global symmetries are imposed:
\begin{itemize}
    \item  A global $U(1)_R$ symmetry plays a crucial role in shaping the structure of the superpotential $W$, enabling the construction of a realistic inflationary model.
    \item A $Z_2$ symmetry that forbids bilinear terms $\Delta_L^a \bar{\Delta}_L^b$ but allows the quartic combination $\Delta_L^a \bar{\Delta}_L^b \Delta_R \bar{\Delta}_R$, thus avoiding the heavy GUT scale mass for the left-triplets.
\end{itemize}
The charge assignments of the superfields under the imposed symmetries are summarized in Table~\ref{tab:symmetry_charges}. These assignments play a critical role in ensuring proton stability by forbidding suppressing baryon number–violating operators. Specifically, the $B\!-\!L$ gauge symmetry prohibits terms such as $Q_c Q_c Q_c$, $L L L_c$, and $L Q Q_c$, while the $U(1)_R$ symmetry forbids operators like $QQQL$ and $Q_c Q_c Q_c L_c$. Together, the structure of the superpotential and the symmetry constraints effectively suppress all proton decay–inducing processes, even those arising from higher-dimensional operators, thereby keeping the model consistent with current experimental bounds on proton lifetime.
\begin{table}[h]
\centering
\begin{tabular}{|c|cccccccccc|}
\hline
\textbf{Field} & $S$ & $\Delta^a_L$ & $\overline{\Delta}^a_L$ & $\Delta_R$ & $\overline{\Delta}_R$ & $H$ & $L$ & $L_c$ & $Q$ & $Q_c$ \\
\hline
$U(1)_R$ & 2 & 2 & 0 & 0 & 0 & 1 & 0 & 1 & 0 & 1 \\

$Z_2$    & +1 & +1 & $-1$ & +1 & $-1$ & +1 & +1 & +1 & +1 & +1 \\

$U(1)_{B-L}$ & 0 & 2 & -2 & 2 & -2 & 0 & 1 & -1 & 1/3 & -1/3 \\
\hline
\end{tabular}
\caption{\small Charge assignments of superfields under $U(1)_R$,  $Z_2$ and $U(1)_R$ symmetries.}
\label{tab:symmetry_charges}
\end{table}
Furthermore, since the vacuum expectation value $\langle \Delta_R \bar{\Delta}_R \rangle$ breaks $B-L$ by two units, matter parity $(-1)^{3(B-L)}$ remains preserved. As a result, the lightest supersymmetric particle (LSP) emerges as a viable dark matter candidate.

The inclusion of left-handed Higgs triplets $\Delta_L^a$ and $\bar{\Delta}_L^a$ provides a natural origin for the type II seesaw mechanism, allowing the generation of tiny neutrino masses. The interaction terms in Eq.~\eqref{eq:superpotential} involving these triplets also permit the decay of the inflaton into lepton-Higgs final states, a key requirement for successful non-thermal leptogenesis.
The first two terms in the superpotential encapsulate the inflationary sector, while the rest describe the interactions necessary for reheating, leptogenesis, and neutrino mass generation. The inflaton field $\Delta_L$ couples to the right-handed triplets and controls their vacuum expectation values (VEVs) during and after inflation. The Yukawa interactions involving $L$ and $L_c$ generate Majorana masses for neutrinos through type II seesaw mechanisms.

\section{Global SUSY Potential}
The relevant part of the superpotential responsible for realizing tribrid inflation is given by
\begin{equation}
W \supset  S \left( \frac{(\Delta_R \bar{\Delta}_R)^2}{M_S^2} - M_X^2 \right) + \frac{\lambda_{ab}}{M_S} (\Delta_L^a \bar{\Delta}_L^b)(\Delta_R \bar{\Delta}_R).
\end{equation}
The first term involving the singlet field $S$ was previously utilized in \cite{Khalil:2012nd} to realize smooth inflation, where the inflationary dynamics is mainly driven by the singlet field $S$. During this phase, the right-handed triplets $\Delta_R = \bar{\Delta}_R$ remain nearly constant and close to the origin, while the left-handed triplets $\Delta_L$ and $\bar{\Delta}_L$ are stabilized exactly at the origin. In this work, we explore a more intriguing scenario in which inflation is driven by left-handed triplets $\Delta_L = \bar{\Delta}_L$, with all other fields stabilized at zero. Remarkably, as shown below, the same field is also responsible for reheating, generating the baryon asymmetry via leptogenesis, and producing tiny neutrino masses through the type II seesaw mechanism.

The charge generator of the current model is given by 
\begin{equation}
Q = T_{3L} + T_{3R} + \frac{B-L}{2} ,  
\end{equation}
where $T_{3L}$ and $T_{3R}$ are the diagonal generators of $SU(2)_L$ and $SU(2)_R$, respectively. It is convenient to write the triplets superfields, $\Delta_{L/R}$ and $\bar{\Delta}_{L/R}$, as \(2 \times 2\) matrices under the left-right gauge symmetry \(SU(2)_L \times SU(2)_R\):
\begin{equation}
\Delta = \begin{pmatrix}
\frac{\Delta^+}{\sqrt{2}} & \Delta^{++} \\
\Delta^0 & -\frac{\Delta^+}{\sqrt{2}}
\end{pmatrix}, \quad 
\bar{\Delta} = \begin{pmatrix}
\frac{\bar{\Delta}^-}{\sqrt{2}} & \bar{\Delta}^0 \\
\bar{\Delta}^{--} & -\frac{\bar{\Delta}^-}{\sqrt{2}}
\end{pmatrix},
\end{equation}
where \(\Delta_{L/R}^{(0,+,++)}\), \(\bar{\Delta}_{L/R}^{(0,-,--)}\) are three components of the triplet superfields, $\Delta_{L/R}$ and $\bar{\Delta}_{L/R}$, respectively. The gauge invariant expressions involving left-right triplets are written as
\[
\Delta \bar{\Delta} \equiv \text{Tr}[\Delta \, \bar{\Delta}] = \text{Tr} \left[
\begin{pmatrix}
\frac{\Delta^+}{\sqrt{2}} & \Delta^{++} \\
\Delta^0 & -\frac{\Delta^+}{\sqrt{2}}
\end{pmatrix}
\begin{pmatrix}
\frac{\bar{\Delta}^-}{\sqrt{2}} & \bar{\Delta}^0 \\
\bar{\Delta}^{--} & -\frac{\bar{\Delta}^-}{\sqrt{2}}
\end{pmatrix}
\right]
= \Delta^+ \, \bar{\Delta}^- + \Delta^{++} \, \bar{\Delta}^{--} + \Delta^0 \, \bar{\Delta}^0,
\]
With these trace definitions, the superpotential is rewritten as:
\begin{equation}
W = S \left( \frac{(\operatorname{Tr}[\Delta_R \bar{\Delta}_R])^2}{M_S^2} - M_X^2 \right) + \frac{\lambda_{ab}}{M_S} \operatorname{Tr}[\Delta_L^{a} \bar{\Delta}_L^{b}] \operatorname{Tr}[\Delta_R \bar{\Delta}_R].
\end{equation}

The Global SUSY scalar potential \( V \) read as:
\begin{align} \label{eq:ftermpo}
V & =  \left| \frac{(\mathrm{Tr}[\Delta_R \bar{\Delta}_R])^2}{M_S^2} - M_X^2 \right|^2 \nonumber \\
&\quad + \left| 2 S \frac{\mathrm{Tr}[\Delta_R \bar{\Delta}_R]}{M_S^2} + \frac{\lambda_{ab}}{M_S} \mathrm{Tr}[\Delta_L^a \bar{\Delta}_L^b] \right|^2 
   \left[ \mathrm{Tr}[\bar{\Delta}_R \bar{\Delta}_R^\dagger] + \mathrm{Tr}[\Delta_R \Delta_R^\dagger] \right]  \nonumber \\
&\quad + \left| \mathrm{Tr}[\Delta_R \bar{\Delta}_R] \right|^2 
   \left( \frac{\lambda_{ab} \lambda_{ac}^*}{M_S^2} \right) 
   \left[ \mathrm{Tr}[\bar{\Delta}_L^b (\bar{\Delta}_L^c)^\dagger] + \mathrm{Tr}[\Delta_L^b (\Delta_L^c)^\dagger] \right] + D-\text{terms} .
\end{align}
The D-terms from the gauge groups \(SU(2)_L\) and \(SU(2)_R\) impose D-flatness conditions during inflation to ensure the scalar potential remains flat. These conditions are satisfied by setting the field components equal: \(\Delta_L^a = \bar{\Delta}_L^a\) and \(\Delta_R = \bar{\Delta}_R\), for \(a = 1, 2\). This alignment ensures the cancellation of the D-terms associated with the non-Abelian gauge groups.
In the \(U(1)_{B-L}\) sector, the D-term contributions from the triplet fields also cancel due to their opposite charges. Specifically, the fields \(\Delta_{L,R}\) and \(\bar{\Delta}_{L,R}\) carry opposite \(B-L\) quantum numbers, leading to a vanishing total D-term.

The global minimum of the scalar potential is 
\begin{equation*}
    \langle S \rangle = 0, \quad \langle  \Delta_L^{a}  \rangle =\langle \bar{\Delta}_L^{a}\rangle=0, \quad \langle \mathrm{Tr}[\Delta_R \bar{\Delta}_R]\rangle = \langle \Delta_R^0 \, \bar{\Delta}_R^0 \rangle = M_{X}M_{S}=M^2
\end{equation*}
We consider a scenario in which the charged components of the $\Delta_{R}$ fields acquire zero vacuum expectation values (VEV) before the onset of observable inflation. This ensures that the corresponding fields, such as 
$\Delta_{R}^{+}$, $\Delta_{R}^{++}$, and their conjugates, do not participate dynamically during inflation and the subsequent waterfall phase transition.
In contrast, the neutral components of the right-handed triplets, namely \(\Delta_R^0\) and \(\bar{\Delta}_R^0\), are assumed to be stabilized at zero during inflation, due to positive mass-squared terms. However, they acquire non-zero vacuum expectation values (VEVs) after the end of inflation, triggered by the waterfall transition mechanism. On the other hand, the left-handed triplet fields \(\Delta_L^0\) and \(\bar{\Delta}_L^0\) are assumed to remain dynamically active during inflation but settle to zero afterward. As briefly discussed below, this scenario cannot be successfully implemented within a global SUSY framework.

The three-field scalar potential $V$ takes the following form in the D-flat direction 
\begin{equation}
\begin{aligned}\label{ftermpo2}
V &= \left( \frac{\chi^4}{2M_S^2} - M_X^2\right) ^2 +\frac{2\chi^2}{M_S^4} \left(2 \chi^2 s  + \frac{M_S\lambda \phi^2 }{2\sqrt{2} } \right)^2+ 
\frac{\lambda^2 \phi^2 \chi^4}{8 M_S^2} 
\end{aligned}
\end{equation}
where  the normalized fields $\Delta_R^0$, $\Delta_L^0$ and $S$ are defined as
\begin{equation*}
    \Delta_R^0=\frac{\chi}{\sqrt{2}},\quad \Delta_L^0=\frac{\phi}{\sqrt{2}},\quad  S=\frac{s}{\sqrt{2}},
\end{equation*}
and the supersymmetric minima shown above have the following form
\begin{equation}
    s=0,\quad \phi=0, \chi=\pm\left(2M_S^2 M_X^2\right)^{1/4}.
\end{equation}
To begin with, successful implementation of tribrid inflation requires the stabilization of the field $s$. However, within the global SUSY framework, it is not possible to generate a sufficiently large mass for $s$, preventing it from settling rapidly at its minimum. Even if we assume $s$ is somehow stabilized, the remaining fields $\chi$ and $\phi$ must then realize a waterfall transition. But this transition fails to occur because the squared mass of the waterfall field $\chi$ at $\chi = 0$, given by $m_\chi^2 = \frac{\lambda^2 \phi^4}{2 M_S^2}$, remains strictly positive, regardless of the value of $\phi$. As a result, the necessary instability to end inflation is absent. To resolve both of these issues and realize a viable tribrid inflation scenario, we must move beyond the global SUSY setup. As we will show below, incorporating supergravity corrections addresses both problems and enables a successful implementation of tribrid inflation in agreement with CMB observations.

\section{SUGRA Inflationary Potential}
The F-term supergravity (SUGRA) scalar potential is given by,
\begin{equation}
V=e^{K/m_{P}^{2}}\left[K_{ij}^{-1}D_{z_{i}} W D_{z_{j}^{*}} W^{*} - 3 m_{P}^{-2}|W|^{2}\right],
\end{equation}
with \(z_{i}\) being the bosonic components of the superfields and where we have defined
\begin{equation}
D_{z_{i}} W=\frac{\partial W}{\partial z_{i}}+m_{P}^{-2}\frac{\partial K }{\partial z_{i}}\mathcal{W}\,,\quad K_{ij}=\frac{\partial^{2} K }{\partial z_{i}\partial z_{j}^{*}}.
\end{equation}
In the present paper, we employ the following form of the
K\"ahler potential including relevant non-minimal terms,
\bea
K &= & |S|^2 + \text{Tr}\left[|\Delta_R|^2\right] + \text{Tr}\left[|\bar{\Delta}_R|^2\right] + \text{Tr}\left[|\Delta_L|^2\right] + \text{Tr}\left[|\bar{\Delta}_L|^2\right] \nonumber \\
& + & \kappa_S \frac{|S|^4}{4 m_P^2} + \kappa_{\Delta R} \frac{\left(\text{Tr}\left[|\Delta_R|^2\right]\right)^2}{4 m_P^2} + \kappa_{\bar{\Delta} R} \frac{\left(\text{Tr}\left[|\bar{\Delta}_R|^2\right]\right)^2}{4 m_P^2}   \nonumber \\  
& + & \kappa_{\Delta L} \frac{\left(\text{Tr}\left[|\Delta_L|^2\right]\right)^2}{4 m_P^2} + \kappa_{\bar{\Delta} L} \frac{\left(\text{Tr}\left[|\bar{\Delta}_L|^2\right]\right)^2}{4 m_P^2}   \nonumber \\
& + & \kappa_{S\Delta R} \frac{|S|^2 \text{Tr}\left[|\Delta_R|^2\right]}{m_P^2} + \kappa_{S\bar{\Delta} R} \frac{|S|^2 \text{Tr}\left[|\bar{\Delta}_R|^2\right]}{m_P^2} + \kappa_{S\Delta L} \frac{|S|^2 \text{Tr}\left[|\Delta_L|^2\right]}{m_P^2} \nonumber \\
& + & \kappa_{S\bar{\Delta} L} \frac{|S|^2 \text{Tr}\left[|\bar{\Delta}_L|^2\right]}{m_P^2}+ \kappa_{\Delta R \bar{\Delta} R} \frac{\text{Tr}\left[|\Delta_R|^2\right] \text{Tr}\left[|\bar{\Delta}_R|^2\right]}{m_P^2} \nonumber \\
& + & \kappa_{\Delta R {\Delta}_L} \frac{\text{Tr}\left[|\Delta_R|^2\right] \text{Tr}\left[|{\Delta}_L|^2\right]}{m_P^2} + \kappa_{\Delta R \bar{\Delta}_L} \frac{\text{Tr}\left[|\Delta_R|^2\right] \text{Tr}\left[|\bar{\Delta}_L|^2\right]}{m_P^2} \nonumber \\
& + & \kappa_{\Delta L \bar{\Delta}_L} \frac{\text{Tr}\left[|\Delta_L|^2\right] \text{Tr}\left[|\bar{\Delta}_L|^2\right]}{m_P^2} + \kappa_{\Delta L \bar{\Delta}_R} \frac{\text{Tr}\left[|\Delta_L|^2\right] \text{Tr}\left[|\bar{\Delta}_R|^2\right]}{m_P^2} \nonumber \\
& + & \kappa_{\bar{\Delta}_L \bar{\Delta}_R} \frac{\text{Tr}\left[|\bar{\Delta}_L|^2\right] \text{Tr}\left[|\bar{\Delta}_R|^2\right]}{m_P^2} + \cdots .
\eea  

The scalar potential with sugra corrections becomes,
\bea
V  & = & \left( \left| \frac{\left( \text{Tr}[\Delta_R \bar{\Delta}_R] \right)^2}{M_S^2} - M_X^2 \right|^2 \right. \nonumber \\
& \times & \left( 1 - \kappa_S \frac{|S|^2}{m_P^2} + (1 - \kappa_{S \Delta R}) \frac{\text{Tr}[|\Delta_R|^2]}{m_P^2}
+ (1 - \kappa_{S \bar{\Delta} R}) \frac{\text{Tr}[|\bar{\Delta}_R|^2]}{m_P^2} \right. \nonumber \\
 &+& \left.  (1 - \kappa_{S \Delta L}) \frac{\text{Tr}[|\Delta_L|^2]}{m_P^2}
+ (1 - \kappa_{S \bar{\Delta} L}) \frac{\text{Tr}[|\bar{\Delta}_L|^2]}{m_P^2} \right) \nonumber \\
& + & \left| 2 S \frac{\text{Tr}[\Delta_R \bar{\Delta}_R]}{M_S^2} + \frac{\lambda_{ab}}{M_S} \text{Tr}[\Delta_L^a \bar{\Delta}_L^b] \right|^2 \left( \text{Tr}[\Delta_R^\dagger \Delta_R] + \text{Tr}[\bar{\Delta}_R^\dagger \bar{\Delta}_R] \right) \nonumber \\
& + & \left| \text{Tr}[\Delta_R \bar{\Delta}_R] \right|^2 \left( \frac{\lambda_{ab} \lambda^{*}_{ac}}{M_S^2}
\left[ \text{Tr}[\bar{\Delta}_L^b \Delta_L^c] + \text{Tr}[\Delta_L^b (\bar{\Delta}_L^c)^\dagger] \right] \right).
\label{eq:scalar_potential}
\eea

\begin{figure}[t]
\centering
\includegraphics[width=0.5\textwidth]{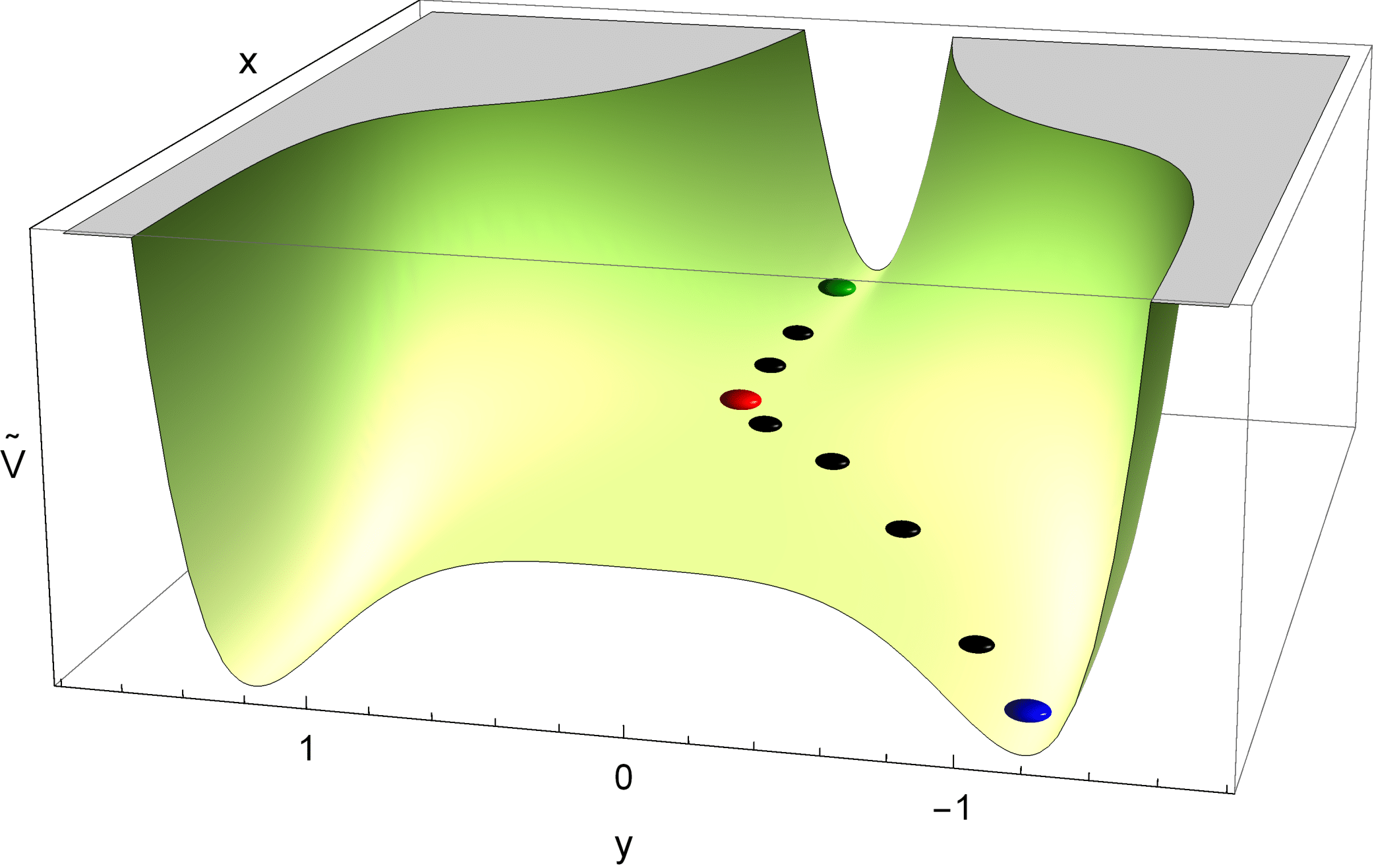}
\caption{
The normalized two-field scalar potential, $\tilde{V} = V/M_X^4$, is shown as a function of the dimensionless variables $x = \phi/M$ and $y = \chi/M$, evaluated along the slice $s = 0$. We assume the critical value $\phi_c = M$, and set the parameter $g = \lambda M / (2 M_X) = 0.1$. The green point marks the onset of observable inflation along the inflationary valley ($\chi = 0$), and the red point indicates the critical (or "waterfall") point, where inflation ends due to the tachyonic instability in the $\chi$ direction. The blue point denotes the global minimum of the potential, where the fields settle after inflation ends and spontaneous symmetry breaking occurs.
}
\label{fig:3Dpot}
\end{figure}

Along the D-flat direction, the scalar potential, involving the three canonically normalized scalar fields ($s$, $\chi$, $\phi$), takes the following form,
\bea \label{ftermpo2}
V (s, \chi, \phi) &=& \left( \frac{\chi^4}{2M_S^2} - M_X^2\right) ^2 +\frac{2\chi^2}{M_S^4} \left(2 \chi^2 s  + \frac{M_S\lambda \phi^2 }{2\sqrt{2} } \right)^2+ 
\frac{\lambda^2 \phi^2 \chi^4}{8 M_S^2}  \nonumber \\
&+& \frac{1}{2} \kappa_S \frac{M_X^4}{m_P^2} s^2 -  \kappa_{\chi} \frac{M_X^4}{m_P^2} \chi^2 +  \kappa_{\phi} \frac{M_X^4}{m_P^2} \phi^2 + \cdots
\eea
with $\kappa_{\chi} = -1 + (\kappa_{S \Delta R} + \kappa_{S \bar{\Delta} R})/2$ and $\kappa_{\phi} = 1 - ( \kappa_{S \Delta L} + \kappa_{S \bar{\Delta} L})/2$.
The first mass term in the second line gives the singlet field $s$ a Hubble-scale mass, $m_s^2 \simeq - 3 \, \kappa_S \mathcal{H}^2$, assuming $\kappa_S \lesssim -1/3$ with $\mathcal{H} \sim M_X^4/3 m_P^2$. This ensures that $s$ rapidly settles to its minimum at the origin, $s = 0$, before the observable inflation. The second mass term contributes to the effective squared mass of the waterfall field $\chi$ at $\chi = 0$ as follows: 
\be
m_\chi^2 = \frac{\lambda^2 \phi^4}{4 M_S^2} - \kappa_\chi \frac{M_X^4}{m_P^2} ,
\ee
which governs the onset of the waterfall transition.
The critical value of the inflaton field $\phi$ at which the waterfall field $\chi$ becomes tachyonic is given by
\be
\phi_c = \left( \frac{4 M_S^2 \kappa_\chi M_X^4}{\lambda^2 m_P^2} \right)^{1/4} = \left( \frac{4 \kappa_\chi M_X^2}{\lambda^2 m_P^2} \right)^{1/4} M.
\ee
Once $\phi$ drops below $\phi_c$, $\chi$ becomes tachyonic and rolls rapidly to its true vacuum, ending inflation. 
To visualize the normalized two-field scalar potential $\tilde{V} = V/M_X^4$ along the slice $s = 0$, we consider the approximate expression:
\be \label{2pot}
\tilde{V}(x,y) \simeq \left( \frac{y^4}{2} - 1 \right) ^2 + g^2  \left( y^2 x^4 + \frac{1}{2} y^4 x^2 - y^2 \right) + \cdots
\ee
where the dimensionless fields are defined as $x = \phi/M$, $y = \chi/M$, and the parameter $g = \lambda M / (2 M_X)$, with the critical value $\phi_c = M$ assumed. The ellipsis denotes additional terms that generate the slope along the inflationary trajectory ($\chi=0$) necessary for slow-roll inflation. This potential is illustrated in Fig.~\ref{fig:3Dpot}, highlighting the inflationary valley along $\chi = 0$, the waterfall transition near $\phi = \phi_c$, and the emergence of the waterfall-Higgs potential around $\phi = 0$.

For $\phi \gg \phi_c$, the waterfall field $\chi$ acquires a large mass and rapidly settles to the origin. In this regime, the inflationary dynamics reduce to an effective single-field description governed by $\phi$. The scalar potential, incorporating the relevant SUGRA corrections is given by
\be  \label{infpot}
V \simeq M_X^4 \left( 1 + \kappa_{\phi} \frac{\phi^2}{m_P^2} + \delta_{\phi} \frac{\phi^4}{m_P^4} \right) .
\ee
Here, the SUGRA-induced terms proportional to $\phi^2$ and $\phi^4$ play a crucial role in generating the required slope for slow-roll inflation. The coefficient $\delta_\phi$ is determined by the K\"ahler potential and is given by
\bea 
\delta_{\phi} &=& \frac{1}{2} + \frac{1}{4} \kappa_{S \Delta_L}^2 + \frac{1}{2} \kappa_{S \Delta_L} \left( -1 + \kappa_{S \bar{\Delta}_L} \right) - \frac{1}{2} \kappa_{S \bar{\Delta}_L} + \frac{1}{4} \kappa_{S \bar{\Delta}_L}^2 , \nonumber \\
&+& \frac{1}{16} \left( \kappa_{\Delta_L} + \kappa_{\bar{\Delta}_L} \right) + \frac{1}{4} \kappa_{\Delta_L \bar{\Delta}_L}. 
\eea
These corrections lift the otherwise flat potential and provide a required slope, allowing for successful slow-roll inflation consistent with observational constraints. 
It is important to note that, in principle, both $\Delta_L^1$ and $\Delta_L^2$ can take part in the inflationary dynamics. However, for simplicity, we consider a scenario in which $\Delta_L^2$ is stabilized at the origin during inflation, and only $\Delta_L^1$ drives the inflationary evolution.

Up to this point, we have worked with the tree-level scalar potential, neglecting one-loop radiative corrections. We now verify that this approximation is justified within the relevant region of parameter space. The components of the $\Delta_R$ and $\bar{\Delta}_R$ superfields acquire Dirac fermion masses $m_F$ and bosonic superpartner masses $m_B$, $m_{\bar{B}}$, with $\phi$-dependent tree-level values given by:
\be
m_F^2  \simeq   \lambda^2 \frac{\Delta_0^4}{M_S^2} , \quad
m_{B}^2  \simeq  \lambda^2 \frac{\Delta_0^4}{M_S^2}  + (1 - \kappa_{S \Delta R})  \frac{M_X^4}{m_P^2}, \quad  m_{\bar{B}}^2  \simeq  \lambda^2 \frac{\Delta_0^4}{M_S^2}  + (1 - \kappa_{S \bar{\Delta} R})  \frac{M_X^4}{m_P^2}.
\ee
where $\Delta_0 = \phi/\sqrt{2}$. The additional terms in the bosonic masses, of order $\mathcal{O}(M_X^4/m_P^2)$, arise from Planck-suppressed K\"ahler corrections and induce a small mass splitting between scalar and fermionic components. During inflation, the one-loop correction to the effective potential takes the form:
\begin{equation}
V_{\text{loop}}  = \frac{3\times 2}{64 \, \Delta^2} \sum\limits_{i} (-1)^{2s_i} m_i^4 \left[  \ln\left( \frac{m_i^2}{Q^2} \right) - 3/2  \right] \simeq \frac{3(\kappa_{S \Delta R} + \kappa_{S \bar{\Delta} R})}{16 \pi^2}  \frac{\lambda^2\phi^4}{M_S^2} \frac{M_X^4}{m_P^2} \ln \left( \frac{\lambda^2\Delta_0^4}{e M_S^2 Q^2} \right),
\end{equation}
where $m_i = (m_B, m_{\bar{B}}, m_F)$ and $s_i = (0,0,1/2)$ denote the mass eigenvalues and corresponding spins of the fields, respectively, and $Q$ is the renormalization scale in the $\overline{MS}$ scheme. Due to the loop suppression factor and the smallness of $\lambda \lesssim 0.01$, as well as the assumed sub-Planckian dynamics, these radiative corrections remain negligible compared to the tree-level potential, confirming the validity of our earlier approximation.

The contribution of soft SUSY-breaking terms, typically at the TeV scale, is generally suppressed in tribrid inflation scenarios, as discussed in \cite{Antusch:2012jc}. In our case, the superpotential $W$ vanishes during the inflationary phase, rendering the soft SUSY-breaking $A$-term negligible. Additionally, the soft scalar mass term $m_{\text{soft}}^2 (\text{Tr}[|\Delta_L|^2] + \text{Tr}[|\bar{\Delta}_L|^2])$ can be safely ignored in comparison to the quadratic term in the inflationary potential, provided that $m_{\text{soft}} \ll \sqrt{\kappa_{\phi}}  (M_X^2/m_P)$. For a typical choice of $M_X \sim 10^{15}$ GeV and TeV-scale soft masses, this condition translates to $\kappa_{\phi}\, \gg 10^{-18}$. This inequality is satisfied in our numerical analysis, justifying the omission of soft SUSY-breaking effects in the inflationary dynamics.

\section{Slowroll Predictions of CMB Observables}
In the slow-roll approximation, the key inflationary observables, the scalar spectral index $n_s$, the tensor-to-scalar ratio $r$, and the running of the spectral index $\alpha_s \equiv dn_s/d\ln k$, are expressed to leading order in the slow-roll parameters as,
\begin{align} \label{rns}
n_s&\simeq 1-6\,\epsilon+2\,\eta\,\,\,, \,\,\, r\simeq 16\,\epsilon, \notag\\
\alpha_{s} &\simeq 16\,\epsilon\,\eta
-24\,\epsilon^2 - 2\,\zeta^2.
\end{align}
The so-called slow-roll parameters, $\epsilon$, $\eta$ and $\zeta$, are defined in terms of the inflationary potential $V$ as,
\begin{align}
\epsilon &= \frac{1}{4} m_P^2
\left( \frac{V'}{V}\right)^2, \,\,\,
\eta = \frac{1}{2} m_P^2 
\left( \frac{V''}{V} \right), \notag\\
\zeta^2 &= \frac{1}{4} m_P^4
\left( \frac{V' V'''}{V^2}\right),
\label{slowroll}
\end{align}
where primes denote derivatives with respect to the field $\phi$ which is related to the canonically normalized inflaton field $\sigma = \sqrt{2}\phi$, and $m_P$ is the reduced Planck mass.
The amplitude of the scalar power spectrum, evaluated at the pivot scale $k_0 = 0.05 \, \text{Mpc}^{-1}$, is given by,
\begin{align}\label{pteq}
A_{s}(k_0) = \frac{1}{24\,\pi^2\,\epsilon}
\left( \frac{V}{m_P^4}\right)\bigg|_{\phi = \phi_0},
\end{align}
where $\phi_0 \equiv \phi(k_0)$ is the field value when the pivot scale exits the horizon. The observed value of the amplitude from Planck is $A_s(k_0) \approx 2.137 \times 10^{-9}$ \cite{Planck:2018vyg, Planck:2018jri}.
The duration of observable inflation is characterized by the number of e-folds $N_0$, given by
\begin{align}\label{Ngen}
N_0 = \frac{2}{m_P^2} \int_{\phi_e}^{\phi_{0}}\left( \frac{V}{%
	V'}\right) d\phi,
\end{align}
where $\phi_e$ is the field value at the end of inflation. In present inflationary scenarios, this is determined by the waterfall transition, $\phi_e = \phi_c$, as defined earlier.

\section{Numerical Analysis and Discussion}
To explore the phenomenological viability of our model and identify regions of parameter space consistent with observational data, we carry out a numerical scan over five key input parameters:
$$
M_X < m_P, \quad M_S \leq m_P, \quad M < \phi_0 \leq m_P, \quad 0 < \kappa_{\phi} < 1, \quad -1 < \delta_{\phi} \leq 0.
$$
The last two conditions correspond to the hilltop inflation regime, which is known to support scenarios with a sizable tensor-to-scalar ratio $r$~\cite{Rehman:2010wm}. 
For the supermassive scale $M_S$, we adopt the perspective that the Planck mass sets the fundamental cutoff of the model. Accordingly, we define $M_S = m_P / \kappa$, with a realistic range $0.1 \leq \kappa \leq 1$.

All accepted solutions are required to satisfy the following observational and theoretical constraints:
\begin{itemize}
\item The amplitude of the scalar power spectrum:
    \[
    A_s(k_0) = 2.137 \times 10^{-9},
    \]
    as specified in Eq.~\eqref{pteq}.   
    
\item  The scalar spectral index, consistent with the central value reported by the ACT collaboration in conjunction with Planck 2018 and LB-BK18~\cite{ACT:2025tim},
\[
n_s = 0.9734.
\]    
\item The number of e-folds, $N_0$, which connects the inflationary dynamics to the thermal history of the Universe. Assuming standard post-inflationary reheating, it is estimated as~\cite{1990eaun.book.....K, Liddle:2003as}:
\begin{align}\label{efolds}
N_0 \simeq 53+\dfrac{1}{3}\ln\left[\dfrac{T_R}{10^9 \text{ GeV}}\right]+\dfrac{2}{3}\ln\left[\dfrac{M_X}{10^{15}\text{ GeV}}\right],
\end{align}
where $T_R$ is the reheating temperature. Details of reheating and inflaton decay are discussed in the next section.
\item The observed baryon asymmetry of the Universe, which is related to the generated lepton asymmetry via sphaleron processes. This requires:
    \[
    \left\vert\frac{n_L}{s} \right\vert \simeq 2 \times 10^{-10},
    \]
as estimated in Eq.~(\ref{nls}).
\end{itemize}
The outcomes of our numerical scan are summarized in Table~\ref{tab:vertical_6bp}, which presents four benchmark points spanning the viable regions of parameter space. Special attention is given to scenarios with an appreciable tensor-to-scalar ratio $r$, including those within the projected sensitivity of next-generation Cosmic Microwave Background (CMB) experiments such as LiteBIRD~\cite{LiteBIRD:2022cnt,LiteBIRD:2024wix}, CMB-S4~\cite{CMB-S4:2020lpa}, and the Simons Observatory~\cite{SimonsObservatory:2018koc}.
The symmetry-breaking scale $M = 2 \times 10^{16}~\text{GeV}$ aligns well with the typical grand unification scale suggested by MSSM gauge coupling unification. Notably, solutions with large $r \gtrsim 0.005$, potentially detectable in future observations, tend to favor slightly higher values such as $M = 5 \times 10^{16}~\text{GeV}$. Furthermore, the running of the spectral index remains within acceptable bounds, with $\alpha_s \lesssim 0.01$ across the viable parameter space.

To gain a deeper understanding of the numerical results, we now examine approximate analytical expressions for the key inflationary observables. We begin with the scalar power spectrum, which can be approximated as
\be \label{Asexp}
A_{s} \simeq  \frac{1}{24\,\pi^2}
\left( \frac{ M_X}{m_P}\right)^4\Bigg[ 
\frac{\kappa_{\phi}\, \phi_0}{m_P} + \frac{2 \delta_{\phi} \, \phi_0^3}{m_P^3} \Bigg]^{-2} = 2.137 \times 10^{-9}.
\ee
where $\phi_0$ denotes the initial inflaton field value.
The tensor-to-scalar ratio $r$ is approximated by
\begin{align}
r &\simeq 16 \left[ \kappa_{\phi} \frac{\phi_0}{m_P} + 2\, \delta_{\phi} \frac{\phi_0^3}{m_P^3} \right]^2 = \frac{2}{3\,\pi^2 A_s} \left( \frac{M_X}{m_P}\right)^4 .
\end{align}
The number of e-folds $N_0$, which measures the duration of inflation, can be expressed as
\be
N_0 \simeq  \frac{1}{2\kappa_{\phi}} \ln \left[ \frac{\phi_0^2}{M^2} \left( \frac{\kappa_{\phi} m_P^2 + 2 \delta_{\phi} M^2}{\kappa_{\phi} m_P^2 + 2 \delta_{\phi} \phi_0^2} \right) \right].
\ee
Assuming $\kappa_\phi > 0$ and $\delta_\phi < 0$, the model realizes hilltop inflation, where the inflaton rolls away from a local maximum. In this regime, the spectral index $n_s$ is dominated by the slow-roll parameter $\eta$, and can be approximated as
\begin{align}
n_s &\simeq 1 + 2 \kappa_{\phi} + 12 \delta_{\phi} \frac{\phi_0^2}{m_P^2}.
\end{align}
The running of the spectral index, $\alpha_s = dn_s/d\ln k$, is given by
\be
\alpha_{s}  \simeq  - \Bigg[ 
\frac{\kappa_{\phi}\, \phi_0}{m_P} + \frac{2 \delta_{\phi} \, \phi_0^3}{m_P^3} \Bigg] \Bigg(  24 \, \delta_{\phi} \frac{\phi_0}{m_P} \Bigg)
.
\ee
These expressions highlight how the inflationary predictions depend sensitively on the interplay between the quadratic term $\kappa_\phi$, the quartic correction $\delta_\phi$, and the initial inflaton field value $\phi_0$. In particular, the quartic contribution introduces important modifications that can either enhance or suppress the tensor-to-scalar ratio $r$, thereby influencing the prospects for observing primordial gravitational waves in future CMB experiments.

\begin{table}[htbp]
\centering
\renewcommand{\arraystretch}{2.2}
\setlength{\tabcolsep}{12pt}
\scriptsize
\begin{tabular}{|l|c|c|c|c|}
\hline
\rowcolor{gray!20}
\textbf{Parameter} & \textbf{BP1} & \textbf{BP2} & \textbf{BP3} & \textbf{BP4}  \\
\hline
\( M~[\text{GeV}] \)         & \( 5.0 \times10^{16} \) & \( 4.5 \times10^{16} \) & \( 3.0 \times 10^{16} \) & \( 2.0 \times 10^{16} \)  \\
\( M_s/m_P \)       & \( 1.2 \times10^{-1} \) & \( 1.1 \times 10^{-1} \) & \( 1.1 \times 10^{-1} \) & \( 9.0 \times 10^{-2} \) \\
\( \phi_0~[\text{GeV}]\)                    & $1.2 \times 10^{18}$ & $6.0 \times 10^{17}$ & $2.5 \times 10^{17}$ & $8.6 \times 10^{17}$  \\
\( M_X~[\text{GeV}] \)       & \( 8.9 \times 10^{15} \) & \( 6.0 \times 10^{15} \) & \( 3.3 \times 10^{15} \) & \( 1.8 \times 10^{15} \)  \\
\( \phi_e/M \)                    & 1 & 1 & 1 & 1\\
\( n_s \)                    & 0.9743 & 0.9743 & 0.9743 & 0.9743 \\
\( r \)                    & $5\times 10^{-3}$ &$1.1\times 10^{-3}$ & $1\times 10^{-4}$ & $8.6\times 10^{-6}$  \\
\( \alpha_s \)    & 0.01 & 0.009 & 0.006 & 0.004  \\
\( N_{0} \)                      & 54.5 & 54.5 & 54.5 & 54.5 \\
\( \kappa_{\phi} \)                  & $0.063$ & $0.059$ & $0.0438$ & $0.0344$  \\
\( \delta_{\phi} \)                      & $-0.054$ & $-0.219583$ & $-0.929$ &$ -5.05927$  \\
\( M_R~[\text{GeV}] \)       & \( 4.5 \times 10^{15} \) & \( 2.6 \times 10^{15} \) & \( 1.1 \times 10^{15} \) & \( 4.6 \times 10^{14} \)  \\
\( T_R~[\text{GeV}] \)       & \( 1.1 \times 10^{9} \) & \( 2.5 \times 10^{9} \) & \( 8.0 \times 10^{9} \) & \( 2.7 \times 10^{10} \)  \\
\( M_1\approx M_2~[\text{GeV}] \)       & \( 7.6 \times 10^{9} \) & \( 1.1 \times10^{10} \) & \( 4.0 \times10^{10} \) & \( 1.2 \times 10^{11} \) \\
\hline
\end{tabular}
\caption{Benchmark points satisfying inflationary and leptogenesis constraints. All mass parameters are in GeV unless otherwise specified.}
\label{tab:vertical_6bp}
\end{table}

For instance, choosing the inflaton field value close to the Planck scale, such as $\phi_0 = m_P/2$, yields a large tensor-to-scalar ratio of $r \simeq 0.005$, as demonstrated by the benchmark point BP1 in Table~\ref{tab:vertical_6bp}. These large-$r$ solutions are central to our numerical investigation. In this regime, the inflationary observables simplify to:
\bea
n_s  & \simeq & 1 + 2 \kappa_{\phi} + 3 \delta_{\phi}, \quad
r \simeq 4 \left[ \kappa_{\phi} + \frac{\delta_{\phi}}{2} \right]^2 = \frac{2}{3\,\pi^2 A_s} \left( \frac{M_X}{m_P}\right)^4 , \\
\alpha_s & \simeq &  - 6 \, \delta_{\phi} \Bigg[ 
\kappa_{\phi} + \frac{\delta_{\phi}}{2} \Bigg] ,  \quad  N_0 \simeq  \frac{1}{2\kappa_{\phi}} \ln \left[ \frac{m_P^2}{M^2} \left( \frac{\kappa_{\phi} }{4\kappa_{\phi} + 2\delta_{\phi} } \right) \right].
\eea
Based on the benchmark point BP2 in Table~\ref{tab:vertical_6bp}, with $\phi_0 = m_P/2$ and $M = 2 \times 10^{16}~\text{GeV}$, which corresponds to a tensor-to-scalar ratio of $r \simeq 0.001$, we can deduce the following parameter values:
\be
\kappa_\phi \simeq 0.06, \quad
\delta_\phi \simeq -0.05, \quad
M_X \simeq 8.7 \times 10^{15}~\text{GeV}, \quad
\alpha_s \simeq 0.01, \quad N_0 \simeq 55.1 \, .
\ee
These values demonstrate excellent agreement with our numerical results, confirming the internal consistency of the model. More generally, for sub-Planckian inflaton values $\phi_0 < m_P$, our parameter scans validate that combinations satisfying the above analytical expressions lead to predictions for inflationary observables that are in good agreement with current data. Furthermore, the model admits the possibility of a sizable tensor-to-scalar ratio, which is potentially within the reach of upcoming CMB polarization experiments.

\section{Inflaton Decay and Reheating}

Let us now discuss inflaton decay and reheating. After inflation, there is a phase of damped oscillation about the supersymmetric vacuum. The oscillating system consists of two scalar fields $\theta_L = (\delta \theta_L + \delta
{\bar\theta}_L)/{\sqrt{2}}$ ($\delta \theta_L = \D^0_L$ and $\delta {\bar\theta}_L = \bar \D^0_L$) and $\theta_R = (\delta \theta_R + \delta
{\bar\theta}_R)/{\sqrt{2}}$ ($\delta \theta_R = \D^0_R - M$ and $\delta {\bar\theta}_R = \bar \D^0_R - M$) with respective masses,
\begin{equation}
M_a^{L} = \lambda_a M \frac{M}{M_S}, \quad \text{ and }
 \quad M_{R} = 2 \sqrt{2} \kappa M \frac{M^2}{M^2_S}.
\end{equation}
Given that $\lambda_a < \sqrt{2} \kappa (M/M_S)$,  $\Delta_R^0$ can decay into a pair of fermionic left triplets  (${\tilde{\D}}_L, {\tilde{{\bar {\D}}}}_L$) respectively through the Lagrangian,
\begin{equation}
L_{\theta_R} =  \sqrt{2} \lambda_a \frac{M}{M_S}
\theta_R {\tilde{\D}^a_L} {\tilde{{\bar {\D}}}^a_L} +
 h.c. .
\end{equation}
The decay width of $\theta_R$ turns out to be%
\be
\Gamma_{\theta_R} = \sum_a {\frac{3}{4\pi}} \lambda^2_a \Bigl
({\frac{M}{M_S}}\Bigr )^2
 M_{R}.
\label{decay}
\ee %
Let us assume the situation where $M_1^L > M_2^L$ as this splitting between $M^L_1$ and $M^L_2$ is important in estimating the lepton asymmetry, as discussed in the next section. The waterfall field $\Delta_R^0$ decays  predominantly into the heaviest fermionic left triplet with mass $M_1^L > M^L_2$, which then, in turn, decays into the lepton-slepton pair and Higgs-Higgsino pair via the following terms in superpotential, 
\begin{equation}\label{decayW}
W \supset  f \Delta_L LL + \frac{\gamma}{M_S} H H \bar{\Delta}_L  \langle \Delta_R \rangle,
\end{equation}
where the index $a=1,2$ is suppressed throughout the following discussion unless explicitly stated otherwise. Note that the decay of $\Delta_R^0$ into right-handed neutrinos is kinematically forbidden since the latter have superheavy mass acquired from the renormalizable coupling $f_R L_c L_c \D_R$, with $f_R$ of order unity.

The other oscillating field, $\Delta_L^{0}$, decays into left-handed neutrinos (sneutrinos) and Higgsinos (Higgses) via the interactions,
\bea
L_{int} &\supset&  - f_{ij}  L_{i} \Delta_L L_{j} - \gamma \frac{M}{M_S}  \tilde{H_d} \bar{\Delta}_L \tilde{H_d} - f_{ij} \lambda \frac{M^2}{M_S}  \tilde{L}^*_{i} \bar{\Delta}_L \tilde{L}^*_{j} -  \gamma \lambda \frac{M^3}{M_S^2}  H_d^* \Delta_L H_d^* + h.c. , \label{Lint} \\
  & \supset & -f_{ij} \Delta^0_L \nu_{i} \nu_{j} -  
 \gamma \frac{M}{M_S} \Delta_L^0 \tilde{H}^0_d \tilde{H}^0_d - f_{ij} \lambda \frac{M^2}{M_S}  \Delta_L^0 \tilde{\nu}^*_{i} \tilde{\nu}^*_{j} -  \gamma \lambda \frac{M^3}{M_S^2} \Delta_L^0 (H_d^0)^*  (H_d^0)^* + h.c.,
\eea
where $\nu$ ($\tilde{\nu}$) are left-handed neutrino (sneutrino) fields, and $H_u (\tilde{H}_u), \, H_d (\tilde{H}_d)$ are electroweak Higgs (Higgsino) doublets of MSSM. It is evident that the decay width of $\Delta_L^0$ to sneutrinos and Higgses is suppressed compared to the decay widths of neutrinos and Higgsinos, which are given by
\begin{equation}
\Gamma(\Delta_L^0 \to \nu \nu) \simeq \sum_{i, j} \frac{1}{16 \pi} M_{L} |f_{ij}|^2, \quad 
\Gamma(\Delta_L^0 \to \tilde{H}^0_d \tilde{H}^0_d) \simeq \frac{\gamma_1^2 \, M^2}{32 \pi \, M_S^2} \, M_L.
\end{equation}
Furthermore, for $|f_{ij}| \gg \frac{\gamma \, M}{4 \sqrt{2\pi} \, M_S} \ll 1$, the decay width of neutrinos becomes the dominant one,
\be
\Gamma_{\theta_L} \simeq  \sum_{i , j} \frac{1}{16\pi} M_{L} |f_{ij}|^2.
\ee
Comparing the decay widths of two oscillating fields, while assuming $|f_{ij}| \lesssim 1$, we get
\be
\frac{\Gamma_{\theta_L}}{\Gamma_{\theta_R}} \simeq  \frac{1}{12 \lambda^2} \frac{M_L}{M_R}  \Bigl
({\frac{M_S}{M}}\Bigr )^2  \gg 1, 
\ee
with $M_R \gtrsim 2 M_L$, $M_S \simeq 100 M$ and $\lambda <  0.01$. This implies that $\Delta_L$-particles decay quickly within one Hubble time and any radiation energy density produced by their decays will be strongly diluted during the matter-dominated phase governed by the oscillations of the $\Delta_R$ field. Therefore, shortly after the inflation, the matter density is dominated by the  $\Delta_R$ field, and it plays the dominant role in the subsequent reheating phase.
The reheating temperature $T_R$ is thus defined in terms of the decay width $\Gamma_{\theta_R}$ as:
\begin{equation}\label{eq:tr}
T_R = \left(\dfrac{90}{\pi^{2}g_{*}}\right)^{1/4}\left(\Gamma_{\theta_R} \, m_{P}\right)^{1/2} \simeq 0.12 \, \lambda \, \frac{M^2}{M_S^2}\sqrt{M m_P} ,
\end{equation}
where $g_{*}= 228.75$ for MSSM.
Consequently, a relatively low reheating temperature, $T_R \lesssim 10^9$ GeV, can be achieved while satisfying the upper bound $\lambda \lesssim \mathcal{O}(10^{-3})$ with $M/M_S \lesssim \mathcal{O}(10^{-2})$ and $M \sim 2 \times 10^{16}$~GeV. Such a reheating temperature remains cosmologically safe, provided the gravitino is sufficiently heavy \cite{Kawasaki:2008qe,Kawasaki:2017bqm,Eberl:2020fml,Lazarides:2020zof,Afzal:2022vjx}. Based on the discussion above, we conclude that by the end of inflation, the oscillating system has fully decayed into $SU(2)_L$ triplet superfields. In the following section, we demonstrate how the subsequent decays of these triplets generate a lepton asymmetry, which is then partially converted into the observed baryon asymmetry via electroweak sphaleron processes \cite{Kuzmin:1985mm,Arnold:1987mh}.

\section{Type II Non-thermal leptogenesis and Neutrino masses}
In general, both right-handed neutrinos and left-handed Higgs triplets can contribute to lepton asymmetry in left-right symmetric models. However, in our scenario where the right-handed neutrinos are superheavy ($M\sim 10^{16}$~GeV), the dominant source of lepton asymmetry arises from the decays of the $SU(2)_L$ triplet fields. 
We mainly follow the analysis of the refs.~\cite{Hambye:2000ui} and \cite{Khalil:2012nd} to estimate the CP asymmetry. For clarity and completeness, we describe the important relations that are relevant to the present model. As described above in Eq.~\eqref{Lint}, the scalar triplet $\D^a_L$ decays into $L L$ and $H_d H_d$, while $\bar \D^a_L$ decays into $\tilde L
\tilde L$ and $\tilde H_d \tilde H_d$. For charged components of the scalar triplets, the relevant interaction terms are 
\bea
L_{int}   &\supset&  \sqrt{2} f_{ij} \Delta^+_L \nu_{i} l^-_{j} + f_{ij} \Delta^{++}_L l^-_{i} l^-_{j} + \sqrt{2} f_{ij} M_L  \bar{\Delta}_L^{-} (\tilde{\nu}_{i})^* (\tilde{l}^{-}_{j})^* +  f_{ij} M_L  \bar{\Delta}_L^{--} (\tilde{l}^{-}_{i})^* (\tilde{l}^{-}_{j})^* + h.c.  \nonumber \\
&+&    \sqrt{2} \alpha \bar{\Delta}_L^- \tilde{H}^0_d \tilde{H}^+_d + \alpha \bar{\Delta}_L^{--} \tilde{H}^+_d \tilde{H}^+_d  + \sqrt{2} \alpha M_L \Delta_L^+ (H_d^0)^*  (H_d^+)^*  + \alpha M_L \Delta_L^{++} (H_d^+)^*  (H_d^+)^* + h.c.,
\eea
where $\alpha = \gamma M/M_S$. Similarly, the respective terms for the fermionic components of the triplet scalar superfields can be written. 


\begin{figure}[t]
\centering
\includegraphics[width=0.75\textwidth]{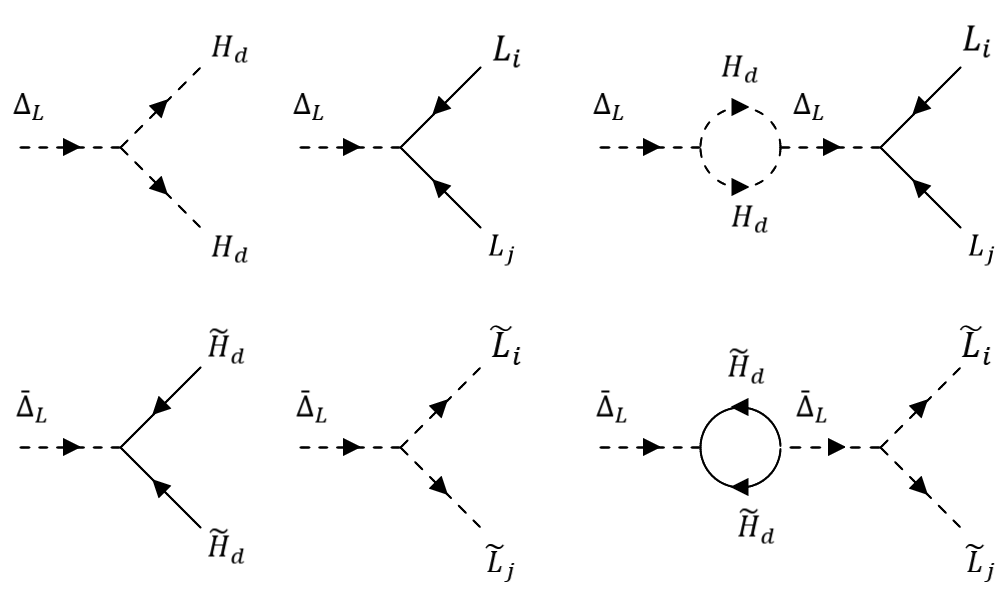}
\caption{
Tree-level and one-loop diagrams contributing to CP asymmetry in the decay of $\Delta_L$ (upper-panel) and $\bar{\Delta}_L$ (lower-panel). Analogous diagrams can also be drawn for their supersymmetric partners.}
\label{fig:combined}
\end{figure}

Tree-level diagrams alone are insufficient to account for CP violation; instead, it arises from the interference between tree-level and one-loop processes (see Fig.~\ref{fig:combined}), as discussed in \cite{Hambye:2000ui}. In leptogenesis scenarios involving triplet Higgs scalars, CP violation originates specifically from self-energy diagrams, where resonant oscillations between nearly degenerate triplet Higgs superfields can significantly enhance the resulting lepton asymmetry. Supersymmetry ensures equal contributions to CP violation from scalar and fermionic decay loops; hence, we focus on scalar triplet decays, noting that their fermionic superpartners yield the same lepton asymmetry.

To compute the CP violation arising from the interference between tree-level and one-loop processes within the mass matrix formalism, we consider the one-loop corrected effective mass terms for the scalar triplets given by:
\be
\Delta_L^{a \dagger} ({\cal M}^2)_{ab} \Delta_L^b +
\bar{\Delta}_L^{a \dagger} (\overline{{\cal M}}^2)_{ab} \bar{\Delta}_L^b,
\ee
where the mass matrices take the form:
\be
{\mathcal{M}}^2 = \begin{pmatrix}
(M_1^L)^2 - i \Gamma_{11} M_1^L & -i \Gamma_{12} M_2^L \\
-i \Gamma_{21} M_1^L & (M_2^L)^2 - i \Gamma_{22} M_2^L
\end{pmatrix},   \quad  \overline{{\mathcal{M}}}^2 = \begin{pmatrix}
(M_1^L)2 - i \bar{\Gamma}_{11} M_1^L & -i \bar{\Gamma}_{12} M_2^L \\
-i \bar{\Gamma}_{21} M_1^L & (M_2^L)^2 - i \bar{\Gamma}_{22} M_2^L.
\end{pmatrix}.  \label{massmatrix}
\ee
The contributions to $\Gamma_{ab}$ ($\bar{\Gamma}_{ab}$) come from the absorptive part of the one-loop self-energy diagrams for $\D^a_L \rightarrow \D^b_L$
($\bar \D^a_L \rightarrow \bar \D^b_L$),
\bea
8\pi \Gamma_{ab} M_b^L &=&  
\sum_{ij} f^{a *}_{ij} f^b_{ij} \, p^2_{\Delta} +
M_a^L M_b^L \, \alpha^a \alpha^{b *} , \nonumber  \\
8\pi \bar{\Gamma}_{ab} M_b^L &=& 
M_a^L M_b^L \sum_{ij} f^a_{ij} f^{b *}_{ij} +
\alpha^{a *} \alpha^b \, p^2_{\Delta} ,
\eea
where $p^2_{\Delta}$ is the squared momentum of the incoming or outgoing particle.

The physical states $\chi^a_{\pm}$ and $\bar{\chi}^a_{\pm}$ are obtained by diagonalizing the mass-squared matrices $\mathcal{M}^2$ and $\overline{\mathcal{M}}^2$. The expressions for these states are given by:
\bea
\begin{pmatrix}
\chi^1_+ \\
\chi^2_+
\end{pmatrix}
&=&
\begin{pmatrix}
1 & -i \Gamma_{12} \, \delta_{2}  \\
i \Gamma_{12}^* \, \delta_{2} & 1
\end{pmatrix} \begin{pmatrix}
\Delta_L^1 \\
\Delta_L^2
\end{pmatrix},
\quad
\begin{pmatrix}
\chi^1_- \\
\chi^2_-
\end{pmatrix}
=
\begin{pmatrix}
1 & -i \Gamma_{12}^* \, \delta_{2} \\
i \Gamma_{12} \, \delta_{2} & 1
\end{pmatrix} 
\begin{pmatrix}
\Delta_L^{1*} \\
\Delta_L^{2*}
\end{pmatrix},  \\
\begin{pmatrix}
\bar{\chi}^1_+ \\
\bar{\chi}^2_+
\end{pmatrix}
&=&
\begin{pmatrix}
1 & -i \bar{\Gamma}_{12} \, \delta_{2}  \\
i \bar{\Gamma}_{12}^* \, \delta_{2} & 1
\end{pmatrix}
\begin{pmatrix}
\bar{\Delta}_L^1 \\
\bar{\Delta}_L^2
\end{pmatrix}, 
\quad
\begin{pmatrix}
\bar{\chi}^1_- \\
\bar{\chi}^2_-
\end{pmatrix}
=
\begin{pmatrix}
1 & -i \bar{\Gamma}_{12}^* \, \delta_{2} \\
i \bar{\Gamma}_{12} \, \delta_{2} & 1
\end{pmatrix} 
\begin{pmatrix}
\bar{\Delta}_L^{1*} \\
\bar{\Delta}_L^{2*}
\end{pmatrix},
\eea
where $\delta_{a} = \dfrac{M_a}{M_1^2 - M_2^2}$. The masses of the physical states are $M_{\chi^a, \bar{\chi}^a} \simeq M_a$, neglecting terms of order $\left(\Gamma_{ab} \delta_{b} \right)^2$. By expressing the scalar fields $\Delta_L^a$ and $\bar{\Delta}_L^a$ in terms of the physical states $\chi^a_{\pm}$ and $\bar{\chi}^a_{\pm}$, and substituting into Eq.~\ref{Lint}, we can compute the decay rates that contribute to the CP asymmetries. Following the definitions in \cite{Hambye:2000ui}, the CP asymmetry for the scalar triplet decays is given by:
\be
\epsilon^a = \Delta L \, \frac{\Gamma(\chi^a_- \rightarrow LL) - \Gamma(\chi^a_+ \rightarrow L^c L^c)}{\Gamma_{\chi^a_-} + \Gamma_{\chi^a_+}} \simeq \frac{M_1 M_2}{2\pi(M_1^2 - M_2^2)}  \frac{\sum_{ij} \text{Im}[f^1_{ij} f^{2*}_{ij} \, \alpha^1 \alpha^{2*}]}{\sum_{ij} |f^a_{ij}|^2 + |\alpha^a|^2},
\ee
and similarly, for the conjugate fields:
\be
\bar{\epsilon}^a = \Delta L \, \frac{\Gamma(\bar{\chi}^a_+ \rightarrow LL) - \Gamma(\bar{\chi}^a_- \rightarrow L^c L^c)}{\Gamma_{\bar{\chi}^a_+} + \Gamma_{\bar{\chi}^a_-}} \simeq \frac{M_1 M_2}{2\pi(M_1^2 - M_2^2)}  \frac{\sum_{ij} \text{Im}[f^1_{ij} f^{2*}_{ij} \, \alpha^1 \alpha^{2*}]}{\sum_{ij} |f^a_{ij}|^2 + |\alpha^a|^2} = \epsilon^a.
\ee
Here, $\Delta L = 2$ reflects the change in the lepton number associated with these decays.

The lepton asymmetry $\left(\frac{n_L}{s}\right)$ generated via type-II leptogenesis in our supersymmetric $SU(2)_L \times SU(2)_R \times U(1)_{B-L}$ framework is given by:

\be
\frac{n_L}{s} = \frac{n_{\Delta}}{s} \sum_a 3\left[\epsilon^a + \bar{\epsilon}^a\right] \simeq \frac{3}{2} \frac{T_R}{M_R} \sum_a 3\left[\epsilon^a + \bar{\epsilon}^a\right],
\ee
where $\frac{n_{\Delta}}{s} \simeq \frac{3}{2} \frac{T_R}{M_R}$ denotes the ratio of the number density of scalar triplets ($n_{\Delta}$) to the entropy density ($s$). Substituting the CP asymmetry expressions, we get:
\be\label{nl_sim}
\frac{n_L}{s} = \sum_a \frac{9}{2}  \frac{T_R}{M_R}  \frac{M_1 M_2}{\pi(M_1^2 - M_2^2)}   \frac{\sum_{ij} \text{Im}[f^1_{ij} f^{2*}_{ij} \, \alpha^1 \alpha^{2*}]}{\sum_{ij} |f^a_{ij}|^2 + |\alpha^a|^2}.
\ee
Once this asymmetry is generated, it's crucial to ensure it survives until electroweak sphaleron processes can convert part of it into baryon asymmetry. In our model, lepton number–violating interactions such as
\be
\Delta_L \leftrightarrow LL, \quad \text{and} \quad \bar{\Delta}_L \leftrightarrow \tilde{L} \tilde{L}, \ \tilde{H}_d \tilde{H}_d \label{e}
\ee
could, in principle, wash out the asymmetry.
However, these wash-out effects are negligible as long as the triplet masses $M_a^L$ are significantly larger than the reheat temperature $T_R$. In our setup, this condition is satisfied for the specific parameter choices, where the ratio $T_R / M_a^L \sim 0.4 \cdot \sqrt{M M_P} / M_S$ remains sufficiently small. This ensures that the generated lepton asymmetry is not erased, making this mechanism more robust than standard thermal leptogenesis.
To estimate the lepton asymmetry $n_L/s$, we first need to fix the parameters appearing in Eq.~\ref{e}, many of which are also relevant to the light neutrino mass matrix.

The neutrino mass matrix is given by the type-II seesaw relation:
\be%
(m_{\nu})_{ij} = 2{f_{ij}} v_{\D_L} - m^T_D m^{-1}_R m_D \equiv
m_{\nu_{II}} - m_{\nu_I},%
\ee%
where $v_{\D_L}$ are the $SU(2)_L$ triplet Higgs's vevs and are obtained from the vanishing of the $F$-term for $\bar{\Delta}_L$ as follows: 
\be
F_{\bar{\Delta}_L} =  \frac{\lambda}{M_S} \Delta_L \langle \Delta_R \bar{\Delta}_R \rangle + \frac{\gamma}{M_S} H H \langle \bar{\Delta}_R \rangle = 0 \quad \rightarrow  \quad
v_{\D_L} = - \frac{\gamma}{\lambda}  \frac{v^2}{M} = -\alpha \frac{v^2}{M_L} . 
\ee
With the masses of all right-handed neutrinos, $m_R = \beta M$, comparable to $M$, $m_{\nu_I}$ are too small to account for the solar and atmospheric neutrino data. Hence, $m_{\nu_{II}}$ provides the main contribution to the neutrino mass matrix, namely 
\be (m_{\nu})_{ij} \simeq - 2
{f^a_{ij}} {\alpha^a  \frac{v^2}{M^a_L}} , \label{mnu} \ee 
where $v \simeq 174$ GeV.

To estimate both the lepton asymmetry
(Eq.(\ref{e})) and neutrino masses (through Eq.(\ref{mnu})), we first
simplify by assuming $|\alpha^1| \simeq |\alpha^2| = \alpha, |f^1| \simeq
|f^2| = f $. Then Eqs.(\ref{e}) and
(\ref{mnu}) can be expressed as 
\bea \frac{n_L}{s} & \simeq &
\frac{9}{\pi} \frac{T_R}{M_{R}}  \times {\frac{\bar{M}}{M_1 - M_2}}
\times {\frac{\Sigma_{ij} |{f_{ij}}|^2 \alpha^2}{\Sigma_{ij} |{f_{ij}}|^2 + \alpha^2}}, 
\\
(m_{\nu})_{ij} &\simeq &2 {f_{ij}} \alpha \frac{v^2}{\bar{M}}, \quad \bar{M} = \frac{M_1 M_2}{M_1 + M_2}, \label{exp} 
\eea 
where we have assumed the CP-violating phase to be maximal.
Expressing $f_{ij} \alpha$ in terms of $(m_\nu)_{ij}$
we obtain
\be\frac{n_L}{s} \simeq \frac{9}{\pi} \frac{T_R}{M_R} \, \frac{\bar{M}}{M_1 - M_2}
\, \frac{\left( \frac{\bar{M}}{2 v^2} \right)^2 \sum_{ij} |(m_\nu)_{ij}|^2 \alpha^2}
{\left( \frac{\bar{M}}{2 v^2} \right)^2 \sum_{ij} |(m_\nu)_{ij}|^2 + \alpha^4}.
\ee
The neutrino mass matrix $m_{\nu}$ can be diagonalized by the PMNS matrix \be
m_{\nu} = U^*_{\nu} {m^{diag}_{\nu}} U^{\dag}_{\nu}, \label{mnur}
\ee where ${m^{diag}_{\nu}}$ = diag($m_{1}, m_{2},
m_{3}$) represent the eigenvalues of the neutrino mass matrix $m_{\nu}$. Using the unitarity of the PMNS matrix and ignoring Majorana phases, we obtain
\be
\sum_{ij} |(m_\nu)_{ij}|^2 = \sum_k m_k^2 .
\ee
This gives:
\be \label{nls}
\frac{n_L}{s} \simeq \frac{9}{\pi} \frac{T_R}{M_R} \, \frac{\bar{M}}{M_1 - M_2}
\, \frac{\left( \frac{\bar{M}}{2 v^2} \right)^2 \sum_k m_k^2 \alpha^2}
{\left( \frac{\bar{M}}{2 v^2} \right)^2 \sum_k m_k^2 + \alpha^4}.
\ee
From global fits \cite{Esteban:2024eli}, the neutrino mass-squared differences are $\Delta m_{21}^2 \equiv m_2^2 - m_1^2 = 7.53 \times 10^{-5}\,\text{eV}^2$, 
$\Delta m_{32}^2 \equiv m_3^2 - m_2^2 = 2.437 \times 10^{-3}\,\text{eV}^2$ (normal hierarchy), and $\Delta m_{32}^2  = - 2.519 \times 10^{-3}\,\text{eV}^2$ (inverted hierarchy).
Assuming normal ordering ($m_1 < m_2 < m_3$) with the minimal allowed $m_1\approx 0$, we obtain $m_2 \approx \sqrt{\Delta m_{32}^2} \approx 8.7\,\text{meV}$ and $m_3 \approx \sqrt{\Delta m_{31}^2} \approx 49\,\text{meV}$, so that $\sum_k m_k^2 = 2.48 \times 10^{-3}\,\text{eV}^2$.
For inverted ordering ($m_3 < m_1 < m_2$) with $m_3 \approx 0$, we obtain
$m_1 \approx m_2 \approx \sqrt{|\Delta m_{32}^2|} \approx 50\,\text{meV}$, so that $\sum_k m_k^2 = 4.88 \times 10^{-3}\,\text{eV}^2$. 
The experimental value of the  baryon-to-photon ratio is given by
\cite{ParticleDataGroup:2024cfk}
\be%
\frac{n_B}{n_{\gamma}} \simeq (6.12 \pm 0.04) \times 10^{-10}.%
\ee%
This baryon asymmetry originates from an initial lepton asymmetry, which is partially converted into baryon number via electroweak sphaleron processes \cite{Kuzmin:1985mm,Arnold:1987mh}. As a result, the required lepton asymmetry is estimated to be:
\be%
\left\vert\frac{n_L}{s} \right\vert \simeq \frac{2}{7} \frac{n_B}{n_{\gamma}} \simeq (1.73 - 1.75) \times 10^{-10}.
\ee

The benchmark points presented in Table~\ref{tab:vertical_6bp} successfully reproduce the above-quoted value of the lepton asymmetry $n_L/s$. These solutions also satisfy the kinematic condition $M_R \geq 2 M_1$, ensuring that the decay of $\Delta_R$ into two $\Delta_L$ fields is allowed. Notably, all benchmark points exhibit an almost degenerate mass spectrum with $M_1 \simeq M_2$, while the mass of the right-handed triplet remains significantly larger, $M_R \gg M_1$.
Crucially, for all viable configurations, the ratio $M_1/T_R \sim 10$, which is sufficient to suppress washout processes that could erase the generated lepton asymmetry. In addition, the reheating temperature remains below the critical threshold for gravitino overproduction, assuming a gravitino mass in the range $m_{3/2} \simeq 10\text{–}50~\text{TeV}$~\cite{Khlopov:1984pf,Ellis:1984eq,Kawasaki:2008qe,Kawasaki:2017bqm}.
These findings provide valuable constraints on the inflaton decay channels and reheating dynamics. Altogether, the analysis highlights the model’s capacity to consistently link inflationary predictions with low-scale leptogenesis, offering a unified and coherent picture of early universe cosmology and the matter-antimatter asymmetry.

\section{Summary and Conclusion}\label{con}
In this work, we have explored a successful implementation of tribrid inflation in a left-right symmetric model based on the gauge group $G_{3221}$. Inflation is driven by the neutral components of the left-handed Higgs triplets, which not only steer the early universe dynamics but also naturally link to the origin of neutrino masses and the generation of the matter--antimatter asymmetry. Neutrino masses are generated through a type-II seesaw mechanism involving right-handed Higgs triplets, which aligns well with current neutrino oscillation data. After inflation, the decay of the inflaton into left-handed triplets enables a successful non-thermal leptogenesis scenario. The resulting lepton asymmetry is converted to a baryon asymmetry via sphaleron processes, in agreement with the observed value.

Supergravity corrections play a crucial role in maintaining the flatness and stability of the inflationary potential. With these effects included, the model yields a scalar spectral index $n_s = 0.9734$, which is well within the range preferred by recent CMB observations, including data from the Atacama Cosmology Telescope. The tensor-to-scalar ratio is predicted to be
$r \lesssim 0.005$, a value that may be probed by next-generation experiments targeting CMB B-mode polarization, such as LiteBIRD and the Simons Observatory.

\section{Acknowledgement}
SOA and MUR extend sincere gratitude to the Deanship of Scientific Research at the Islamic University of Madinah, Saudi Arabia, for the support provided.

\bibliographystyle{apsrev4-1}

\bibliographystyle{unsrt}  
\bibliography{References}  

\end{document}